\documentclass[12pt]{article}

\usepackage{graphpap}
\usepackage{graphics}

\begin{document}

\setlength{\baselineskip}{4ex}

\title{{\em A Priori} Approach to Calculation of Energies of Solvation}
\author{Sohrab Ismail-Beigi$^\dag$\\
$\dag${\em Department of Physics, Massachusetts Insitute of Technology},\\
{\em Cambridge, MA 02139},\\
\\
T.A.~Arias$^\ddag$\\
$\ddag${\em Laboratory of Atomic and Solid State Physics, Cornell
University,}\\ {\em Ithaca, NY 14853}\\
$\ddag${\em Research Laboratory of Electronics, Massachusetts Institute
of Technology,}\\ {\em Cambridge, MA 02139},\\
\\
Phillip Marrone, Matthew Reagan, and Jefferson W. Tester$^*$\\
$^*$ {\em Department of Chemical Engineering and Energy Laboratory},\\
{\em Massachusetts Institute of Technology},\\
{\em Cambridge, MA  02139}\\
}
\date{\today}
\maketitle
\vspace{0.5cm}

\begin{abstract}
We propose a systematic, {\em a priori} approach to the problem of the
calculation of solvation energies using continuum dielectric models
coupled to quantum mechanical description of reacting molecules.  Our
method does not rely on empirically scaled van der Waals radii to
create a dielectric cavity, but rather uses the electron density of
the reactants as the physical variable describing the cavity.  In
addition, the precise choice of cavity is made by ensuring that the
dielectric reproduces the correct linear response of the solvent to
electrostatic perturbations.  As a model application which is
interesting in its own right, we study the hydrolysis of methylene
chloride, a representative model waste compound in supercritical
oxidation experiments, and which has shown surprising solvation
effects close to the critical point of pure water (T=394$^o$ C, P=221
bar).  Using our {\em a priori} methodology, we find results in good
agreement with available experimental reaction barriers.  We then
study, in a controlled manner, the relative importance of various
further approximations that are routinely performed in the literature
such as the use of spherical cavities, the replacement of the reactant
by a dipole, or the neglect of self consistency in solving the
electrostatic problem.
\end{abstract}

\tableofcontents
\listoftables
\listoffigures

\section{Introduction}

The study of reactions in solution and the calculation of solvation
energies are problems of importance and interest in physics,
chemistry, and biology.  For a polar solvent such as water, the
effects of solvation on the energetics of reactions can be crucial.
However, modeling solvated reactions is a challenging problem as one
needs both a reliable quantum mechanical treatment of the reactants in
order to correctly describe bond rearrangements as well
as thermodynamic integration over the solvent degrees of freedom.

An idealized model would describe both the reactants and the large
number of solvent molecules quantum mechanically using an {\em ab
initio} approach.  However, simulating such a large number of
particles quantum mechanically poses a prohibitive computational
burden even on today's largest computers, and we require some sort of
simplification or approximation scheme.  Clearly, the electronic
states of the reacting molecules must be treated quantum mechanically,
so that one must concentrate on simplifying the description of the
solvent.  The most common and direct approach replaces the solvent by
a dielectric continuum that surrounds a solvent-excluded cavity about
the reactants, and we refer the reader to the excellent review of
Tomasi and Persico \cite{ChemRev}.

Upon examining the state of the art, one sees that current
methodologies are highly variegated and that they rely on
semiempirical fits or rule-of-thumb prescriptions for describing the
dielectric cavity, the charge density of the reactants, or the
polarization of the dielectric.  We instead consider
systematically the impact of each stage of approximation in
coarse-graining from an {\em ab initio} calculation one could perform
in principle to the dielectric treatment used in practice.

In addition to this new perspective, we provide a novel approach for
dealing with the important issue of constructing appropriate
dielectric cavities.  The current literature uses van der Waals
spheres to construct the cavity.  The results, however, can be
sensitive to the radii of the spheres requiring empirical adjustment
\cite{ChemRev,Spheres}.  Rather than this {\em a posteriori} approach,
we construct the cavity based on the {\em a priori} consideration that
the dielectric, by definition, reproduces the thermodynamic response
of the molecular solvent to electrostatic perturbations.  Below, we
effect this by choosing the dielectric cavity boundary as an
isosurface of the reactant electron density which reproduces the
correct solvent response.

Once we specify the dielectric cavity, we still face the problem of
solving the electrostatic equation.  For this, we introduce a powerful
preconditioned conjugate-gradients method exploiting Fourier transform
techniques to solve the electrostatic problem in the presence of the
dielectric cavity.

To explore the efficacy of our approach, we consider a reaction that
is highly sensitive to the dielectric response of the solvent, the
hydrolysis of methylene chloride (CH$_2$Cl$_2$).  There is both
experimental (e.g. \cite{FMH,Phil2,Phil3,MODAR,Sandia}) and
theoretical interest (e.g. \cite{Phil1}) in the aqueous breakdown of
this industrial toxin (methylene chloride).  Previous theoretical
studies of this reaction showed promising results, but were based on
the simple approximation of a dipole in a spherical cavity
\cite{Phil1}.  In addition to providing a more systematic analysis of
the hydrolysis of methylene chloride, we study in a controlled manner
the impact of these common approximations on the resulting solvation
energies.

We begin in Section~\ref{sec:reaction} with a discussion of the
hydrolysis reaction under consideration.  There, we present the
overall hydrolysis reaction for methylene chloride and identify the
rate limiting step, the choice of reaction geometry and reaction
coordinate, and the {\em ab initio} density-functional method used to
treat the reactants.  Next, Section~\ref{sec:theory} describes our
theoretical approach beginning with how one, in principle, calculates
free energies {\em a priori} and then presents a detailed analysis of
the chain of approximations that lead to the dielectric model.  Our
method for creating the dielectric cavity is presented followed by a
new algorithm for solving the electrostatic problem.  Finally, in
Section~\ref{sec:results} we present results for the hydrolysis of
methylene chloride followed by a detailed analysis of the impact of
the dipole and sphere approximations and the importance of
self-consistency.

\section{Pathway for hydrolysis of methylene chloride}
\label{sec:reaction}

Methylene chloride (CH$_2$Cl$_2$) is an important industrial solvent.
Recently, there has been interest in treating aqueous wastes
containing residual CH$_2$Cl$_2$ using supercritical water oxidation
(SCWO) (i.e. oxidation in water at conditions above its critical point
at 374$^o$ C and 221 bar \cite{Phil1}).  However, significant
destruction of CH$_2$Cl$_2$ occurs during preheating at subcritical
conditions through hydrolysis rather than oxidation.  Perhaps more
surprisingly, the hydrolysis reaction rate is found to decrease
dramatically as temperature increases through the critical point
\cite{Phil1,Phil2}, signalling the importance of solvation effects for
this reaction.

The overall reaction describing the hydrolysis of methylene chloride
is
\[
\mbox{CH}_2\mbox{Cl}_2 + \mbox{H}_2\mbox{O} \rightarrow \mbox{HCHO} +
2\mbox{HCl}\,.
\]
This reaction is known to be a two step process \cite{FMH}.  The first
step is a slow, rate-limiting substitution process that dictates the
overall rate of the above reaction,
\begin{equation}
\mbox{CH}_2\mbox{Cl}_2 + \mbox{H}_2\mbox{O} \rightarrow
\mbox{CH}_2\mbox{ClOH} + \mbox{HCl}\,.
\label{firstreaction}
\end{equation}
The species CH$_2$ClOH is unstable and undergoes a fast internal
rearrangement that expels H$^+$ and Cl$^-$ to form the final HCHO,
\[
\mbox{CH}_2\mbox{ClOH} \rightarrow \mbox{HCHO} + \mbox{HCl}\,.
\]
Therefore, we need only concentrate our attention on the rate-limiting
step of Eq.~(\ref{firstreaction}).

The first step of the reaction of Eq.~(\ref{firstreaction}) requires
an H$_2$O molecule to approach the CH$_2$Cl$_2$ molecule very closely
and for a Cl$^-$ ion to leave.  This proceeds via the creation of a
transition state,
\begin{equation}
\mbox{CH}_2\mbox{Cl}_2 + \mbox{H}_2\mbox{O} \rightarrow
\mbox{CH}_2\mbox{ClOH}_2 + \mbox{Cl}^-
\label{reactionequation}
\end{equation}
Below, we concentrate on understanding the energetics of the
transition-state complex of Eq.~(\ref{reactionequation}), which should
provide us with the activational energy barrier of the reaction of
Eq.~(\ref{firstreaction}).

\begin{figure}[t!]
\begin{center}
\resizebox{3.0in}{!}{\includegraphics{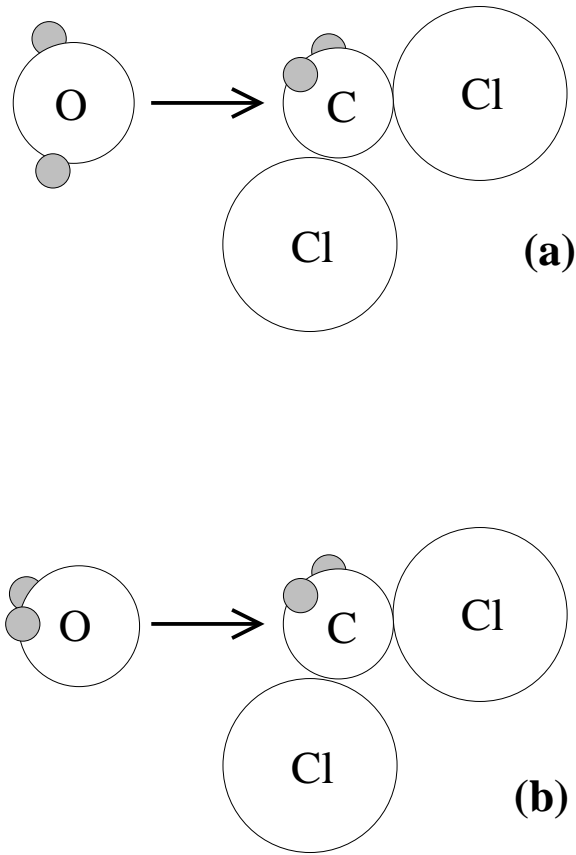}}
\end{center}
\caption[Stereochemistry of hydrolysis reaction] {Schematic diagrams
of the stereochemistry of CH$_2$Cl$_2$ hydrolysis.  Shaded spheres
represent hydrogen atoms and other atoms are labeled.  The arrow shows
the direction of approach of the H$_2$O molecule.  Possibilities (a)
and (b) are discussed in the text.}
\label{fig:reaction}
\end{figure}

The chemistry of the reaction of Eq.~(\ref{reactionequation}) is known
to be of the S$_N$2 variety, where the oxygen approaches the carbon
from the side opposite to the chlorine ion that leaves the
CH$_2$Cl$_2$ molecule.  However, the precise orientation of the water
molecule during this reaction is not known.  Figure \ref{fig:reaction}
shows that the hydrogens in the water molecule may be oriented in one
of two ways, either (a) in the same plane as the carbon and chlorine
atoms, or (b) rotated by $90^o$.  Resolution of this question requires
{\em ab initio} calculations.

First principles calculations provide unambiguous, {\em a priori}
results for energies and forces.  Such {\em ab initio} calculations
determine reactant geometries, energies, and charge densities in
vacuum.  We use the pseudopotential plane-wave density-functional
approach in the local-density approximation \cite{RMP} with the
Perdew-Zunger parameterization \cite{PerdewZunger} of the
Ceperly-Alder exchange-correlation energy \cite{CeperleyAlder}.
Non-local pseudopotentials of the Kleinmann-Bylander form \cite{KB}
constructed using the optimization scheme of Rappe {\em et al.}
\cite{Rappe} describe the interaction of valence electrons with the
ionic cores.  The pseudopotential for carbon has a non-local projector
for the {\em s} channel, and the oxygen and chlorine have projectors
for the {\em p} channel.  The plane-wave cutoff is 40 Rydbergs for a
total of 35,000 coefficients for each electronic wave function.  For a
given choice of ionic positions, electronic minimizations are carried
out using a parallel implementation of the conjugate-gradient
technique of reference \cite{RMP}.  The supercell has dimensions of
15 \AA\ $\times$ 9 \AA\ $\times$ 9 \AA, which separates periodic
images of the reactants sufficiently to minimize spurious interaction
effects, even for the elongated transition states.  We fix the carbon
atom at the origin of our simulation cell and place the reaction
coordinate $\lambda$, defined as the oxygen-carbon distance of the
reactants in Eq.~(\ref{reactionequation}), along the long, 15 \AA\
x-axis of our cell.  We determine optimized molecular structures by
moving the ionic cores along the Hellman-Feynman forces until all
ionic forces (except along the {\em fixed} reaction coordinate
$\lambda$) are less than 0.1 eV/\AA\ in magnitude.  To illustrate the
accuracy of these calculations, Table \ref{table:vacprops} compares
experimental and {\em ab initio} values for bond-lengths and permanent
dipole moments of the isolated molecules.

\begin{table}[t!]
\begin{center}
\begin{tabular}{|ccc|}
\hline
& Bond Lengths (\AA)& \\
\hline
bond\ \ \ \  & {\em ab initio} & experimental \cite{bondlengths} \\
\hline
C-H   &   1.11  &  1.09 \\
C-Cl  &   1.79  &  1.77 \\
O-H   &   0.99  &  0.96 \\
\hline
\hline
& Dipole Moments (Debyes)& \\
\hline
molecule\ \ \  & {\em ab initio} & experimental \cite{dipolemoments} \\
\hline
H$_2$O       & 1.89 & 1.85 \\
CH$_2$Cl$_2$ & 1.79 & 1.6  \\
\hline
\end{tabular}
\end{center}
\caption{{\em Ab initio} bond lengths and dipole moments}
\label{table:vacprops}
\end{table}

The {\em ab initio} calculations establish that  pathway (b) of
Figure~\ref{fig:reaction} is preferred, being 0.21 eV lower in energy
for a typical value of the reaction coordinate $\lambda$.  We thus
consider only this pathway in the remainder of our work.  Towards
this end, we catalogue optimized geometries for a series of values of
reaction coordinate along this pathway.

\section{Theoretical methodology}
\label{sec:theory}

Having determined the energies and configurations along the pathway in
vacuum, we now turn to the much more challenging problem of the
calculation of Gibbs free energies.  Of particular interest is $\Delta
G^*$, the difference between the Gibbs free energy of the solvated
transition state $G^\ddag$ and the Gibbs free energy of the solvated
reactants at infinite separation $G^j$,
\begin{equation}
\Delta G^* \equiv G^\ddag - \sum_{j} G^j\,.
\label{Gstardef}
\end{equation}
Next we define the free energy of solvation $G^i_{solv}$ as the
difference in Gibbs free energy of a configuration $i$ in vacuum
$G^i_0$ (which we have calculated above {\em ab initio}) and in
solution $G^i$,
\begin{equation}
G^i = G^i_0 + G^i_{solv}\,.
\label{Gsolvdef}
\end{equation}
The task is to compute $G^i_{solv}$ which describes the interaction of
the solvent with the reactants.

\subsection{Microscopic treatment of Gibbs free energies}
\label{sec:theoroverview}

Direct calculation of the Gibbs free energy $G^i$ requires the
evaluation of a large phase space integral,
\begin{equation}
e^{-G^i/k_BT} = \int' dq_r \int dq_s\ e^{-\left[H_r(q_r) +
H_{r,s}(q_r,q_s) + H_s(q_s)\right]/k_BT},
\label{Gdef}
\end{equation}
where $q_r$ and $q_s$ are coordinates describing the positions of the
reactant and solvent nuclei, respectively, and $H_r$, $H_s$, and
$H_{r,s}$ are the Hamiltonians describing the isolated reactants,
isolated solvent, and their interaction, respectively.  We work within
the Born-Oppenheimer approximation, and therefore these Hamiltonians
are the corresponding system energies for fixed nuclear coordinates
$(q_r,q_s)$.  Finally, the prime on the outer $q_r$ integral indicates
that we only sum over reactant coordinates that are compatible with
the configuration $i$.

In principle, there is no fundamental difficulty in (a) preparing a
cell containing the reactants and a large collection of solvent
molecules, (b) computing the system energy $H_r + H_s + H_{r,s}$ in
Eq.~(\ref{Gdef}) within density-functional theory, and (c) integrating
over the phase space with appropriate molecular dynamics or Monte
Carlo methods.  The only hindrance is the prohibitive computational
effort required.  Accurate description of the bonding rearrangements
of the chemically active reactants requires quantum mechanical
calculation of the reactant Hamiltonian $H_r$.  Therefore, the only
option to render the computation tractable while maintaining accuracy
is somehow to coarse-grain the detailed microscopic description of the
solvent.

\subsection{Coarse-graining the solvent}
\label{sec:linearresp}

Our approach to coarse-graining the solvent replaces the detailed
molecular arrangement of the solvent molecules by the electrostatic
field $\Phi(r)$ which they generate.  To accomplish this, we separate
the interaction of the reactant charge distribution $\rho_r(r)$ and
the solvent electrostatic field $\Phi$ from all other terms in the
reactant-solvent interaction Hamiltonian,
\begin{equation}
H_{r,s}(q_r,q_s) = \int d^3r\,\rho_r(r)\Phi(r) \ + \ V(q_r,q_s)\,.
\label{Hrsbreakdown}
\end{equation}
The term $V$ includes all remaining reactant-solvent interactions such
as hard-core repulsions and van der Waals attractions (due to induced
dipole dispersion forces).

Next, we define the free energy $G_s[\Phi]$ of the solvent
{\em given} that it produces a field $\Phi(r)$ through
\begin{equation}
e^{-G_s[\Phi]/k_BT} = \int_{q_s\rightarrow\Phi}
dq_s\ e^{-[H_s(q_s)+V(q_r,q_s)]/k_BT},
\label{Gsdef}
\end{equation}
where $q_s\rightarrow\Phi$ means that we only integrate over
configurations $q_s$ that give rise to the field $\Phi$.  The
only dependence of the free energy $G_s$ on the configuration of
the reactants $q_r$ is through the interactions $V$, the most important
being the hard-core repulsions that create the solvent-excluded cavity
about the reactants.  Therefore, $G_s[\Phi]$ is the
coarse-grained description of the free energy of the solvent in the
presence of the cavity.

Next, we expand $G_s$ about its minimum at $\Phi_c$,
\begin{equation}
G_s[\Phi] = G_s[\Phi_c] + {1\over 8\pi}\int d^3r\,
\left(\Phi(r)-\Phi_c(r)\right) \hat{K}
\left(\Phi(r)-\Phi_c(r)\right) + ...\ 
\label{Gsexp}
\end{equation}
Here, $\hat{K}$ is a positive-definite symmetric kernel which depends
on the cavity shape and which specifies the thermodynamic response of
the solvent in the presence of the cavity, and $\Phi_c$ is the
electrostatic field created by the solvent in the presence of an empty
cavity.  As we shall discover below, retaining only the quadratic
expansion of $G_s$ ultimately leads to a familiar dielectric
description.

The free energy $G^i$ of Eq.~(\ref{Gdef}) now can be
rewritten in terms of $G_s$ by integrating over all possible solvent
fields $\Phi$,
\begin{equation}
e^{-G^i/k_BT} = \int' dq_r\,e^{-H_r(q_r)/k_BT} \int d\Phi\
e^{-\left[\int d^3r\, \rho_r(r)\Phi(r) \ + \
G_s[\Phi]\right]/k_BT}\,,
\label{GiwithGs}
\end{equation}
which within the quadratic approximation of Eq.~(\ref{Gsexp}) becomes
\begin{eqnarray}
e^{-G^i/k_BT} & = & \int' dq_r \int d\Phi\ e^{-\left[H_r(q_r) +
\int d^3r\, \rho_r\Phi + G_s[\Phi_c] + {1\over
8\pi}\int d^3r\, \left(\Phi-\Phi_c\right) \hat{K}
\left(\Phi-\Phi_c\right)\right]/k_BT}\nonumber\\ & = & \int'
dq_r\ e^{-\left[H_r(q_r) + {1\over 2}\int d^3r\,
\rho_r\left(\phi_s+\Phi_c\right) + G_c\right]/k_BT}.
\label{Gilinear}
\end{eqnarray}
In going from the first to the second line, we have performed the
Gaussian integral over the solvent field $\Phi$, which introduces two
new quantities.  The variable $\phi_s(r)$ in Eq.~(\ref{Gilinear}) is the
electrostatic potential at the maximum of the Gaussian integrand as
determined by the condition,
\begin{equation}
\hat{K}\left[\phi_s(r)-\Phi_c(r)\right] = -4\pi\rho_r(r)\,,
\label{phisdef}
\end{equation}
from which it is clear that the linear operator $\hat{K}$ relates the
electrostatic response of the solvent $\phi_s(r)$ to the presence of
the reactant charges $\rho_r(r)$.  The constant $G_c$ is the free
energy of formation of the empty cavity.  Physically, the contents of
the square brackets of the exponent in Eq.~(\ref{Gilinear}) represent
the total free energy of the system for a fixed reactant configuration
$q_r$.  This free energy consists of the internal energy of the
reactants $H_r(q_r)$, the electrostatic interaction of the reactants
with the response of the solvent $\int d^3r\, \rho_r\,\phi_s$, the
interaction of the solvent with the cavity potential $\int d^3r\,
\rho_r\,\Phi_c$, and the cavitation free energy $G_c$.

For the case of present interest, the solvation of molecules with
permanent electrical moments, we would expect the induced solvent
potential $\phi_s$ to greatly exceed the cavitation potential
$\Phi_c$, which arises from an electrostatically neutral cavity, and
therefore that the electrostatic reactant-solvent interactions will be
the dominant contribution to the free energy.  Indeed, studies of
polar molecules show the total solvation free energy to be strongly
correlated with the aforementioned electrostatic interaction.
Although these two quantities have an absolute offset of about 0.2~eV,
free energy differences between configurations may be computed to
within 0.05~eV from differences in the electrostatic interaction alone
\cite{ChemRev}.  Mathematically, this implies that we can set
$\Phi_c=0$ and that $G_c$ may be taken to be independent of the
reactant configuration $q_r$, resulting in
\begin{eqnarray}
e^{-G^i/k_BT} & = & e^{-G_c/k_BT} \int' dq_r\ e^{-\left[H_r(q_r) +
{1\over 2}\int d^3r\,
\rho_r(r)\phi_s(r)\right]/k_BT}\,,\label{Gilinearelec}\\
\hat{K}\phi_s(r) & = & -4\pi\rho_r(r)\,.\label{Kelec}
\end{eqnarray}
The free energy of solvation for configuration $q_r$ is the
electrostatic integral
\begin{equation} 
G_{solv} = {1\over 2} \int d^3r\ \rho_r(r)\,\phi_s(r)\,.
\label{Gsolv}
\end{equation}

Eq.~(\ref{Kelec}) shows that the reactant charge density $\rho_r$
induces a linear response in the solvent which gives rise to the
solvent potential $\phi_s$.  This relation, therefore, also gives the
total electrostatic potential $\phi=\phi_s+\phi_r$ as a linear
response to the charge density of the reactants,
\[
\left[\hat{K}^{-1} + \nabla^{-2}\right]^{-1} \phi = - 4 \pi \rho_r(r).
\]
This latter connection corresponds precisely to the standard
macroscopic Maxwell's equation,
\begin{equation}
(\nabla\cdot \epsilon \nabla) \phi = -4\pi\rho_r(r)\,,
\label{PBeq}
\end{equation}
and serves to {\em define} the precise form of $\epsilon$ in terms of
$\hat K$, which we have already defined microscopically.

\subsection{Comparison of continuum and molecular response}
\label{sec:SPC}

Having arrived at a coarse-grained description of the solvent in terms
of its dielectric function in Eq.~(\ref{PBeq}), we now face the
problem of specifying $\epsilon$.  In general, the true microscopic
dielectric is a non-local function which relates to the cavity in a
highly complicated manner.   However, as we now show, quite simple,
{\em computationally tractable} models can describe very well
the underlying physics.

Within the solvent-excluded cavity surrounding the reactant, there is
no solvent dielectric response and therefore we expect $\epsilon=1$.
Far from the cavity, we expect the dielectric response to be that of
the bulk solvent $\epsilon= \epsilon_b(P,T)$.  Here, we include the
dependence of the bulk solvent's dielectric constant $\epsilon_b$ on
pressure $P$ and temperature $T$ as we wish to investigate dielectric
effects near the critical point of water where $\epsilon$ is a strong
function of these parameters.  Finally, we can expect the dielectric
response somehow to interpolate smoothly between these extremes.

To investigate the suitability of such a simple dielectric description
of an ordered molecular solvent, we in principle could use {\em ab
initio} methods by placing a large number of water molecules in a
simulation cell, then performing molecular dynamics or Monte Carlo
sampling, and comparing the response to electrostatic perturbation of
this system with the response of a dielectric model.  This, however,
would be both computationally expensive and largely unnecessary as the
solvent molecules remain chemically inert and interact with the
reactants primarily via electrostatic and repulsive forces.
Therefore, to investigate the impact of coarse-graining the molecular
details of the solvent, we employ a simpler model where the solvent's
electronic degrees of freedom are not described explicitly.  The
microscopic SPC model for water was used as it has electrostatic point
charges for the three atoms in the H$_2$O molecule and a Lennard-Jones
interaction between the oxygen centers to incorporate short-range
repulsive and long-range van der Waals interactions \cite{SPC}.

We extract the linear response of H$_2$O to electrostatic
perturbations by simulating the behavior of SPC water molecules about
a spherical cavity.  We represent the cavity as a Lennard-Jones
potential with a diameter of 4.25 \AA\ and with a well-depth of 0.0030
eV centered at the origin of the simulation cell.  (This corresponds
to $\sigma_{ij}=4.25$ \AA\ and $\varepsilon_{ij}=0.0030$ eV in the
notation of \cite{SPC}.) This cavity has roughly the same size and
radius of curvature as the reactant complex that we study below, and a
small depth parameter is chosen so that the potential primarily
presents a repulsive core to oncoming water molecules.  At the center
of the cavity, we place a point charge of $Q=0,\pm 0.1e,\pm 0.2e$ so
that $\rho_r(r) = Q\delta^3(r)$ in Eq.~(\ref{PBeq}).  For each value
of $Q$, we run constant $(N,P,T)$ molecular dynamics simulations
\cite{NPMD,NTMD} with $N$=256 H$_2$O molecules, $P$=1~atm, and
$T=298$~K, which corresponds to an average cell volume of
7,800~\AA$^3$.  We equilibrate the cell for 25~ps before gathering
data for each time step over a run-time of 250~ps.  The time-averaged
oxygen-cavity radial distribution functions show a large peak at a
radius of 3.0$\pm$0.1 \AA\ and no oxygen presence for smaller radii.
This gives an effective ``hard core'' radius of 3~\AA\ for this
spherical cavity.

\begin{figure}[t!]
\begin{center}
\resizebox{4.0in}{!}{\includegraphics{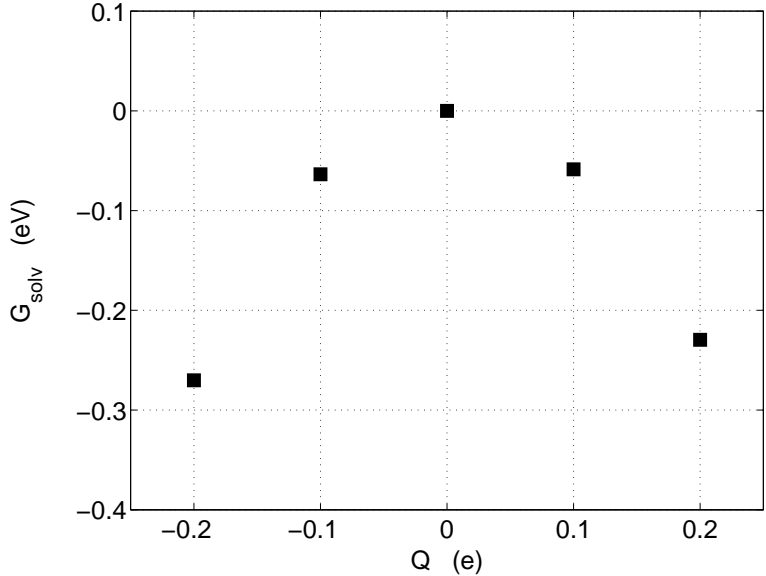}}
\end{center}
\caption[Solvation energy for spherical cavity simulations]
{Electrostatic solvation free energy for the spherical cavity SPC
simulations as a function of $Q$, the strength of the point-charge at
the center of the cavity.}
\label{fig:GsolvversusQ}
\end{figure}

Binning and averaging the instantaneous charge distributions over the
250~ps sampling period gives the average induced charge density in the
solvent $\rho_s$, from which we may calculate $\phi_s$ using
$\nabla^2\phi_s=-4\pi\rho_s$.  We then compute the solvation free
energy of Eq.~(\ref{Gsolv}).  Figure \ref{fig:GsolvversusQ} shows
quadratic behavior in $Q$, the first indication that the dielectric
approach is appropriate.

As discussed above, we now compare this molecular response with that
of a dielectric function which interpolates smoothly from the interior
of the cavity to deep within the solvent,
\begin{equation}
\epsilon(r) = 1 + {(\epsilon_b(P,T)-1) \over 2}\ \mbox{erfc} \left(
{r_\epsilon-r \over \sqrt{2}\sigma}\right)\,,
\label{epsilonformula}
\end{equation}
which is the convolution of a Gaussian having a standard deviation
$\sigma$ with a step function located at a radius of $r_\epsilon$.
This particular form of interpolation is chosen for convenience when
working with our Fourier-based technique described further below.
Here, $r_\epsilon$ is the radius where the dielectric changes and
$\sigma$ measures the distance over which the change occurs.

\begin{figure}[t!]
\begin{center}
\resizebox{3.5in}{!}{\includegraphics{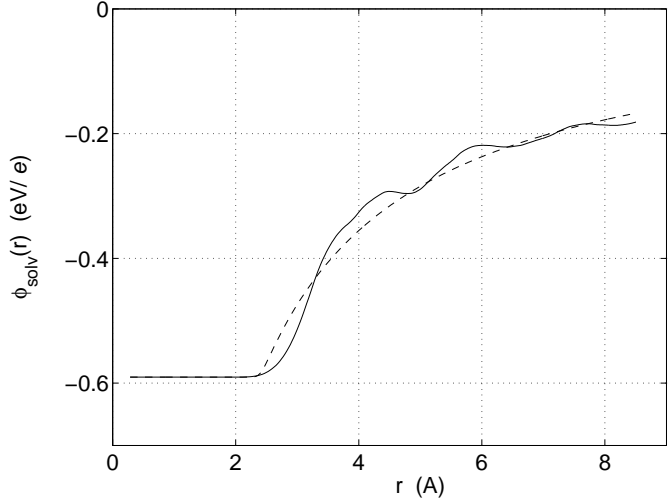}}
\end{center}
\caption[Electrostatic potential for spherical cavity simulations]
{Electrostatic potentials created by the solvent $\phi_s$ for the
spherical cavity versus distance $r$ from the cavity center.  The
solid line is calculated from the time-averaged solvent charge density
of the SPC molecular dynamics simulations. The dashed line is the
potential from the dielectric model. Here, a charge $Q=+0.1e$ was
placed at $r=0$ and the cavity has Lennard-Jones diameter of
4.25~\AA. The dielectric model has $r_\epsilon=2.65$ \AA,
$\sigma=0.11$ \AA, and $\epsilon_b=80$.}
\label{fig:SPCdielectricpot}
\end{figure}

If a continuum dielectric description is viable, then an appropriate
choice of $r_\epsilon$ and $\sigma$ will ensure that, in general, the
response of the model dielectric match that of the molecular solvent
and, in particular, that the electrostatic free energies matches as
closely as possible.  Figure \ref{fig:SPCdielectricpot} shows $\phi_s$
for $Q=+0.1e$ as calculated from the molecular simulations and the
dielectric of Eq.~(\ref{epsilonformula}) with $r_\epsilon=2.65$~\AA,
$\sigma=0.11$~\AA, and $\epsilon_b=80$ as appropriate for water at
ambient conditions.  For this demonstration, $\sigma=0.11$~\AA\ is
fixed and $r_\epsilon$ was chosen to minimize the mean square
difference between the molecular and dielectric potentials $\phi_s$.
Although some shell structure is evident in the molecular calculation,
the two potentials track one another quite closely.  Moreover, because
the system is a point charge, the solvation free energy of
Eq.~(\ref{Gsolv}) may be read off as simply the value of $\phi_s$ at
the origin, where the two calculations agree quite well.

\subsection{Specification of dielectric cavity}
\label{sec:constructepsilon}

As shown above, the dielectric approximation to the response of the
solvent appears quite successful in practice.  Encouraged by this, we
now specify how we construct the dielectric cavity for our reactant
system.  Specifically, a closed surface surrounding the reactants must
be chosen, a surface that specifies the boundary where the dielectric
changes from its value in vacuum $\epsilon=1$ to its value in the bulk
solvent $\epsilon=\epsilon_b(P,T)$.

The current state of the art for choosing the cavity begins by placing
spheres with empirically adjusted van der Walls radii on atomic or
bond sites.  These adjusted radii are scaled from the experimentally
fit radii by factors in the range of 1.15 to 1.20 \cite{ChemRev}.
Next, either the volume enclosed by the intersecting spheres is taken
as the dielectric cavity, or a further spherical probe is rolled over
this volume and either the surface of contact of the probe or the
surface traced by its center is used as the cavity boundary
\cite{ChemRev}.  The latter two choices tend to either underestimate
the solvation energy or to produce unphysical inward-bulging regions
that lead to computational difficulties and unphysical sensitivities
to slight changes in reactant geometry \cite{ChemRev}.  When this
approach is stable, calculated solvation energies can depend strongly
on the size and placement of the spheres \cite{Spheres}.  Aside from
the empirical nature of the approach, it is not obvious why spherical
shapes should be used in molecules where electron densities can differ
significantly from a sum of spherical atomic densities, or why it is
appropriate to use atomic van der Waals radii in a molecule.

Our {\em a priori} approach is based on the principle that the
dielectric cavity captures the thermodynamic response of the solvent
to the charge density of the reactants.  Our strategy for applying
this idea has two parts.  First, we know that strong repulsive forces
between reactant and solvent molecules create the cavity, and that
these repulsive forces are active when the reactant and solvent
electron clouds overlap thereby causing the system's energy to rise
rapidly due to the Pauli exclusion principle.  Therefore, the shape of
the surface of closest approach for the solvents and hence the shape
of the dielectric cavity should be well approximated by isosurfaces of
the electron density of the reactants.  Second, to choose the precise
value of the electron density specifying the isosurface, we demand
that the dielectric so chosen lead to the correct solvation free
energy as predicted by an {\em ab initio} molecular description.  In
practice, this requirement is very difficult to enforce for an
arbitrarily shaped cavity.  Therefore, as a necessary practical
compromise, we instead ensure that (a) our dielectric model produces
the correct molecular response of the solvent for a computationally
manageable model system, and (b) we use relevant {\em ab initio}
calculations to calibrate the results of the model calculations when
applying them to the real system.  We now provide the details below.

\subsubsection{Molecular-dielectric connection}

Our first step is to connect the molecular description to the
dielectric one so as to extract key physical parameters for use in our
{\em a priori} modeling.  To this end, we concentrate on the results
of our SPC-cavity simulations described above.

The first important parameter is the radius $r_{O}$, defined as the
position of the first maximum in the cavity-oxygen radial distribution
function and therefore the closest-approach distance of the oxygen
atoms to the cavity center.  As stated above, we have $r_O = 3.0\pm
0.1$~\AA.

The second important parameter is $r_{peak}$, defined to be the
position of the induced charge peak within the dielectric model.  For
a radial dielectric function such as that of
Eq.~(\ref{epsilonformula}), the induced charge density in response to
a point-charge $Q$ at the origin is given by
\begin{equation}
\rho_{ind}(r) = {Q \over 4\pi r^2}{d \over dr}{1 \over 
\epsilon(r)}=-{Q \over 4\pi r^2}{\epsilon'(r) \over \epsilon(r)^2}\,.
\label{rhoind}
\end{equation}
Thus $r_{peak}$ is the radius $r$ where $\rho_{ind}$ has its largest
magnitude.  For each value of $\sigma$, we choose the optimal
dielectric model by finding the $r_\epsilon$ that minimizes the mean
square difference between the $\phi_s$ generated by the SPC
calculation and that of our dielectric model.

Two important results emerge from this fitting.  First, for all values
of $Q$ and $\sigma$, we find $r_{peak}$ to be essentially constant,
$r_{peak}=2.3\pm 0.2$ \AA, which means that the spatial position of
the induced charge distribution remains fixed.  While we expect the
closest-approach distance $r_O$ to depend on the details of the
repulsive potential, the difference $r_O-r_{peak}$, which is
determined primarily by the geometry of the water molecule itself,
should be more transferable.  Therefore, we always will design our
dielectric functions such that $r_O-r_{peak}=0.7$~\AA.  The second
result relates to the choice of $r_\epsilon$, which is the final
parameter necessary to complete the specification of the dielectric
function $\epsilon(r)$.  In order to keep $r_{peak}$ fixed, we find
that $r_\epsilon$ follows the simple relation $r_\epsilon = r_{peak} +
2.3\sigma$ to within $\pm$~0.03~\AA.  This completes the specification
of $\epsilon(r)$ for ambient conditions $\epsilon_b=80$.

For $\epsilon_b\ne 80$, we can, in principle, repeat the above
procedure.  However, $r_O-r_{peak}$ was fixed to keep it independent
of the value of the bulk dielectric $\epsilon_b$ for the reasons
described above.  Then, we determine $r_\epsilon$ such that the peak
position of $\rho_{ind}$ (Eq.~(\ref{rhoind})) falls at the resulting
value of $r_{peak}$.

\subsubsection{Construction of non-spherical dielectric cavities}

The first problem in determining the dielectric cavity for a reacting
complex is to determine the closest approach of the solvent water
molecules.  To address this, we start with an isolated CH$_2$Cl$_2$
molecule and bring an H$_2$O molecule toward the carbon atom from the
direction opposite to one of the chlorine atoms. The incoming molecule
is oriented with its hydrogens pointing away from the carbon so as to
allow for the closest approach of the oxygen.  Then the {\em ab
initio} energy of the system is calculated as a function of the
carbon-oxygen separation and a rapid rise over and above $k_BT$ is
found for a separation of 2.5~\AA.  This provides an {\em ab initio}
closest-approach distance of $r_O=2.5$~\AA.  Using the fixed
separation $r_O-r_{peak}=0.7$~\AA\ of the previous section, we conclude
that $r_{peak} = 1.8$~\AA.

To convert $r_\epsilon$ into a dielectric boundary in three
dimensions, we scan the electron density of an isolated CH$_2$Cl$_2$
along the approach direction of the H$_2$O molecule in the previous
paragraph.  The electron density value $n_\epsilon$ at the distance
$r_\epsilon$ from the carbon is then recorded, and this isosurface of
the electron density defines the boundary of the dielectric cavity.

\subsection{Solving the Poisson equation}
\label{sec:solvePB}

The remainder of this section describes how we solve the Poisson
equation~(\ref{PBeq}) for the total potential $\phi(r)$.  When solving
for $\phi$, one faces two computational issues: (a) how to represent
the continuous fields $\phi$, $\rho_r$, and $\epsilon$, and (b) to
what level of accuracy one solves the Poisson Eq.~(\ref{PBeq}).

We choose to represent all continuous functions with a periodic
Fourier expansion.  In particular, a Fourier grid of dimension
$200\times 120\times 120$ is used for the 15~\AA\ $\times$ 9~\AA\
$\times$ 9~\AA\ cell corresponding to a grid spacing of 0.075~\AA.
The represention of $\rho_r$ and $\epsilon$ on a grid requires us to
address the issue of the discreteness of that grid. The total reactant
charge density $\rho_r$ is the sum of a smooth electronic charge
density and the point-like charge densities of the atomic cores, the
latter presenting the essential difficulty when representing $\rho_r$
on the grid.  The dielectric function $\epsilon$ likewise has rapid
variations across the boundary of the cavity.  We use Gaussian
smoothing to deal with both problems: we replace each atomic core by a
Gaussian distribution, and we smooth the dielectric by convolving it
with the same Gaussian.  The standard deviation $\sigma$ of the
Gaussian smoothing must be both small enough to faithfully recover the
same energy differences that we would obtain with true true point
charges and also larger than the grid spacing so that we can
faithfully represent $\rho_r$ and $\epsilon$.

\begin{figure}[t!]
\begin{center}
\resizebox{4.0in}{!}{\includegraphics{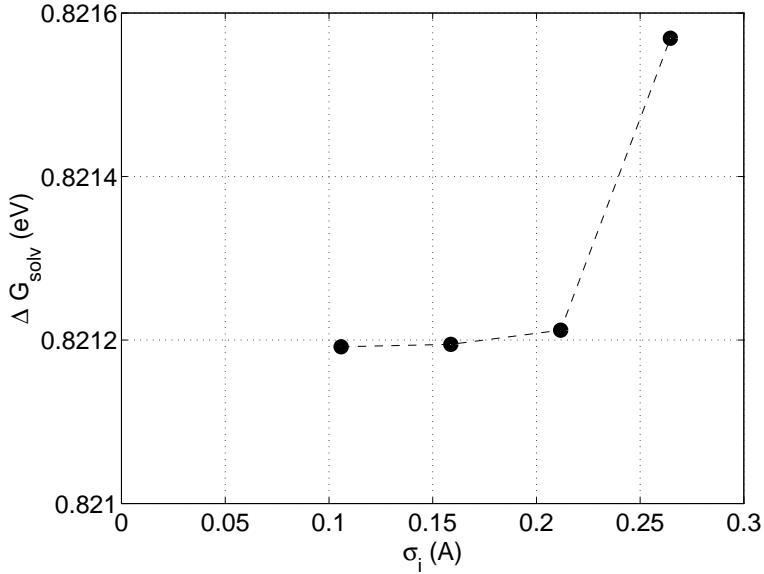}}
\end{center}
\caption[Effect of ionic smoothing on free energy differences]{The
difference of electrostatic solvation free energy between
configuations $\lambda=1.41$ \AA\ and $\lambda=1.84$ \AA\ versus the
ionic smoothing length $\sigma_i$.  The dielectric smoothing was held
fixed at $\sigma_\epsilon=0.11$ \AA.  The dashed curve is a guide for
the eye.}
\label{fig:ionsmooth}
\end{figure}

Our high-resolution grid allows us to satisfy both constraints.
Figure \ref{fig:ionsmooth} shows the behavior of the electrostatic
solvation free energy difference between two reactant configurations
as a function of $\sigma$ for a fixed dielectric.  Choosing $\sigma =
0.11$~\AA\ introduces errors in free energy differences of less than
$10^{-4}$ eV, consequently this value of $\sigma$ was selected for all
calculations reported below.

Since any function can be expanded in a Fourier basis, our choice of
representation is quite general and free of any {\em a priori} bias.
By simply increasing the size and density of the Fourier grid, we are
guaranteed systematic convergence of the results.

Finally, to solve the Poisson equation~(\ref{PBeq}) accurately, 
the auxiliary quadratic functional ${\cal L}[\phi]$ is minimized:
\begin{equation}
{\cal L}[\phi] = {1 \over 8\pi}\int d^3r\, \epsilon(r)
\left|\vec\nabla\phi(r)\right|^2 - \int d^3r\, \rho_r(r)\phi(r)\,,
\label{Ldef}
\end{equation}
which is equivalent to solving the Poisson Eq.~(\ref{PBeq}).  Very
briefly, we represent $\phi(r)$ in our Fourier basis via
\begin{equation}
\phi(r) = \sum_q \hat{\phi}(q)\,e^{i q \cdot r}\,, 
\end{equation}
where $q$ ranges over the Fourier wave-vectors, and $\hat{\phi}(q)$
are Fourier expansion coefficients.  In this representation,
calculation of $\vec\nabla\phi$ is equivalent to to multiplication of
$\hat{\phi}(q)$ by $iq$.  Multiplication by the dielectric function
$\epsilon(r)$ and reactant charge density $\rho_r(r)$ is best
performed in real-space, where we multiply by the value of $\epsilon$
or $\rho_r$ at each grid point.  Integration is replaced by a weighted
sum over grid points in real-space, and we use Fast Fourier transforms
to effect the change of representation from reciprocal to real space
and vice versa.  In this way, ${\cal L}$ becomes a quadratic function
of the Fourier coefficients $\hat\phi(q)$, and minimization of a
quadratic function is an ideal case for the use of conjugate gradient
methods.  Convergence was accelerated by preconditioning each
component of the gradient ${\partial{\cal L}
\over\partial\hat\phi(q)}$ by the multiplicative factor ${4\pi\over
q^2}$ for $q\ne 0$ and unity for $q=0$.  With this choice, the
minimization requires only twenty to twenty five iterations to reach
machine precision.  Once we have found $\phi$, we easily obtain
$\phi_s = \phi - \phi_r$ and calculate the solvation free energy of
Eq.~(\ref{Gsolv}).

\subsection{The rigid solute approximation}
\label{sec:rigidsolute}

Up to this point, we have not dealt with the reactant coordinates
$q_r$.  Even within a dielectric model, performing the integral of
Eq.~(\ref{Gilinearelec}) poses a difficult problem.  This integral
over $q_r$ is approximated by evaluation at its maximum, resulting in
\begin{equation}
G^i = \min_{q_r} \left[ H_r(q_r) + {1 \over 2}\int d^3r\,
\rho_r(r)\phi_s(r)\right]\,.
\label{Gimax}
\end{equation}
Finding $G^i$ poses a tedious self-consistent problem.  The key
difficulty is that the reactant charge density $\rho_r$ and solvent
field $\phi_s$ are implicit functions of $q_r$ so that to perform the
above minimization, one must determine the solution of the quantum
mechanical equations that determine $H_r(q_r)$ while {\em
self-consistently} solving the external Poisson Eq.~(\ref{PBeq}) for
$\phi_s$.

The simplest and most popular approach is to ignore the polarizability
of the reactants and to use their optimal configuration in vacuum.
Therefore, $G^i$ of Eq.~(\ref{Gimax}) becomes simply the sum of the
energy of the reactants in vacuum and the electrostatic interaction of
the resulting reactant charge density with the solvent.  This ``rigid
solute'' approximation, when compared to the full self-consistent
minimization, creates absolute errors in the free energy ranging from
0.26 eV to 0.52 eV, but differences of free energies are generally
accurate to better than 0.03 eV \cite{ChemRev}, which is sufficient
for our study.

\section{Results and Discussion}
\label{sec:results}

\subsection{Reaction profile and barrier}
\label{sec:reacbarrier}

Based on the above methodology, we first perform {\em ab initio}
calculations at a set of reaction coordinates in the range 1.4 \AA\
$<\lambda<$ 2.0 \AA along the the pathway defined for the reaction in
Section~\ref{sec:reaction}.  We also perform one calculation at
$\lambda=7.5$ \AA, the largest separation possible in our cell, which
serves as our reference configuration and which we label as
$\lambda=\infty$ (reactants at essentially infinite separation).  The
{\em ab initio} energy of each configuration is then $G^i_0$ in
Eq.~(\ref{Gsolvdef}).

\begin{figure}[t!]
\begin{center}
\resizebox{4.5in}{!}{\includegraphics{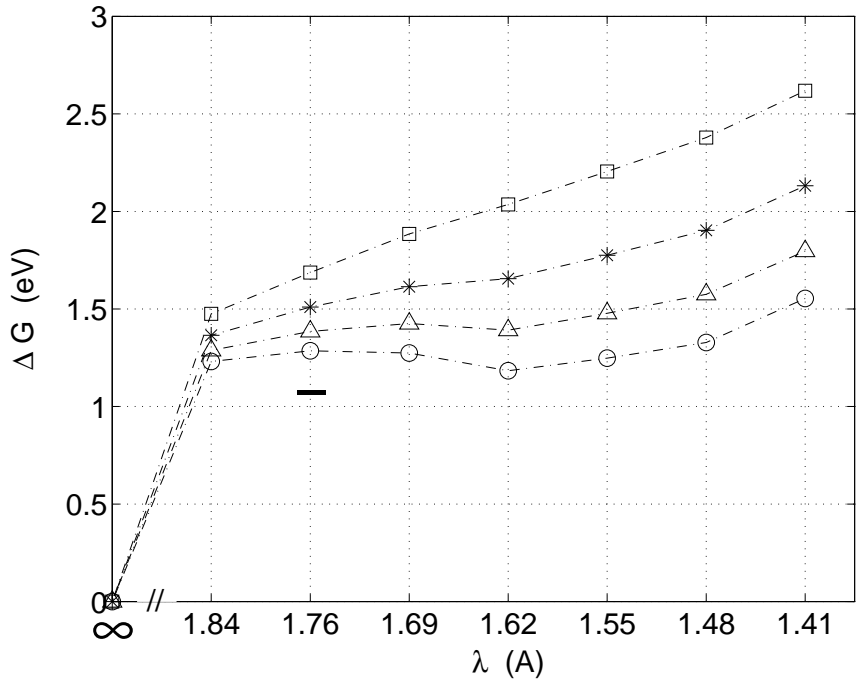}}
\end{center}
\caption[Reaction profile] {Free energies changes $\Delta G$ of the
reacting complex versus reaction coordinate $\lambda$ for various
values of the bulk water dielectric constant $\epsilon_b$.  Circles
are $\epsilon_b=80$, triangles $\epsilon_b=5$, stars $\epsilon_b=2$,
and squares $\epsilon_b=1$.  Free energy changes are relative to the
reactants at infinite separation $\lambda=\infty$.  The solid dash at
1.11~eV represents the experimental value of the reaction barrier at
$T=388 K$ \protect \cite{FMH} where $\epsilon_b=52$ \protect
\cite{UF}.}
\label{fig:abireac}
\end{figure}

Next, using the {\em ab initio} charge density calculated at each
$\lambda$, we create appropriate dielectric cavities for various
values of the bulk dielectric constant $\epsilon_b$ as described
above.  We solve the Poisson equation for each case and calculate the
solvation free energy $G^i_{solv}$.  This solvation energy is added to
$G^i_0$ to give the total free energy of the configuration $G^i$.
Figure \ref{fig:abireac} shows the resulting free energies of the
reacting complex as a function of reaction coordinate $\lambda$
referenced to the $\lambda=\infty$ configuration for various values of
$\epsilon_b$.  The curve with $\epsilon_b=1$ shows the raw {\em ab
initio} energies of each configuration in vacuum $G^i_0$.

The results in Figure \ref{fig:abireac} show the following trends:

(a) For large values of the dielectric constant $\epsilon_b$, the free
energy curve has a local maximum at $\lambda\approx 1.7$ \AA\
followed by a shallow minimum at smaller $\lambda$.  The height of
this maximum is the activation free energy $\Delta G^*$ of
Eq.~(\ref{Gstardef}) for our reaction.

(b) As $\epsilon_b$ decreases, the heights of both the maximum and therefore
also the activation barrier increase.  This is because a weaker dielectric
leads to a weaker induced charge in the solvent, to a smaller
solvation energy, and thus to less stabilization of the reacting complex.

(c) As $\epsilon_b$ decreases, the maximum and minimum move closer and
finally coalesce.  The free energy curve then becomes rather
featureless and rises monotonically for decreasing $\lambda$.

The trends (b) and (c) explain the qualitative behavior of the
hydrolysis reaction of Eq.~(\ref{reactionequation}) observed in
experiments \cite{FMH,Phil2}.  As the temperature $T$ increases, the
dielectric constant of water drops sharply from its ambient value of
$\epsilon_b=80$ to $\epsilon_b\approx 1$.  As $\epsilon_b$ decreases,
the height of the barrier $\Delta G^*$ rises which drastically reduces
the reaction rate.

As for quantitative comparison with experimental values, the work of
Fells and Moelwyn-Hughes \cite{FMH} found an activation barrier of
$\Delta G^* = 1.11\pm0.02$ eV at $T=388 K$, which corresponds to
$\epsilon_b=52$ \cite{UF}.  This experimental value is marked as the
solid dash in Figure \ref{fig:abireac}.  The {\em a priori} results
predict a barrier $\Delta G^*=1.3$ eV.  The local-density
approximation used in the {\em ab initio} calculations is generally
considered to have an absolute accuracy of 0.1-0.2 eV for such a
system.  In addition, the other approximations listed above together
with their appropriate uncertainties introduce a further uncertainty
which we estimate to be about 0.05~eV.  Therefore, our {\em a priori}
prediction of the barrier compares favorably with the experimental
value.

Trend (c) raises the question of the nature of the reaction of
Eq.~(\ref{reactionequation}) for high temperatures or, equivalently,
low dielectric $\epsilon_b$.  A simple interpretation of the data in
Figure \ref{fig:abireac} based on transition state theory would imply
that there is no stable product and that the reaction ceases as
$\epsilon_b$ tends to one.  However, since we have not explored the
entire possible phase space for the two step reaction of
Eq.~(\ref{firstreaction}), it would be na\"{\i}ve to claim that the
reaction stops completely.  More probably, for this high temperature
regime, some other reaction path or mechanism begins to compete with
the one we examine here.  In this case, free radical reaction pathways
become more important as the temperature rises or equivalently
$\epsilon_b$ decreases.

In summary, our {\em a priori} approach explains the behavior of the
hydrolysis reaction Eq.~(\ref{firstreaction}) as a function of
temperature.  Furthermore, the quantitative comparison to available
experimental data is favorable given the approximations involved.

\subsection{Kirkwood theory}
\label{sec:kirk}

The idea of modeling a solvent by a dielectric continuum dates back to
the early 1900's as illustrated in the works of Born \cite{Born} and
Bell \cite{Bell}.  Kirkwood \cite{Kirkwood} developed a picture for
the interaction of a dielectric continuum with a quantum mechanical
solute.  Within Kirkwood's formulation, the solvent-excluded cavity is
a sphere of radius $R$.  The dielectric inside is $\epsilon=1$, and,
outside the sphere, the dielectric takes its bulk value
$\epsilon=\epsilon_b$. One then performs a multipole expansion of the
solute charge density.  Solving for the induced surface charge density
on the spherical boundary is then a standard textbook problem.  The
resulting expression for the electrostatic solvation free energy
$G^i_{solv}$ is
\begin{equation}
G^i_{solv} = {Q_i^2e^2 \over 2R}\left({1 \over \epsilon_b} -1\right) +
{p_i^2 \over R^3}\left({1-\epsilon_b \over 1+2\epsilon_b}\right) + ...
\label{Kirkformula}
\end{equation}
where $Q_ie$ is the total charge of the solute, $\vec p_i$ is its
electrical dipole moment, and contributions of higher moments of the
solute charge distribution are not shown.  Generally, the leading term
of the series dominates and higher order terms can be neglected.  In
our case, we have neutral but polar reactants CH$_2$Cl$_2$ and H$_2$O,
so that the series begins with the dipolar term, the only term
retained.  We compute the dipole moment $\vec p_i$ of the reactant
configuration $i$ directly from the {\em ab initio} charge density.
Then, the one free parameter in the Kirkwood formalism is the cavity
radius $R$.  We choose $R$ such that the volume of the Kirkwood sphere
and our {\em a priori} dielectric cavity are the same.

\begin{figure}[t!]
\begin{center}
\resizebox{4.5in}{!}{\includegraphics{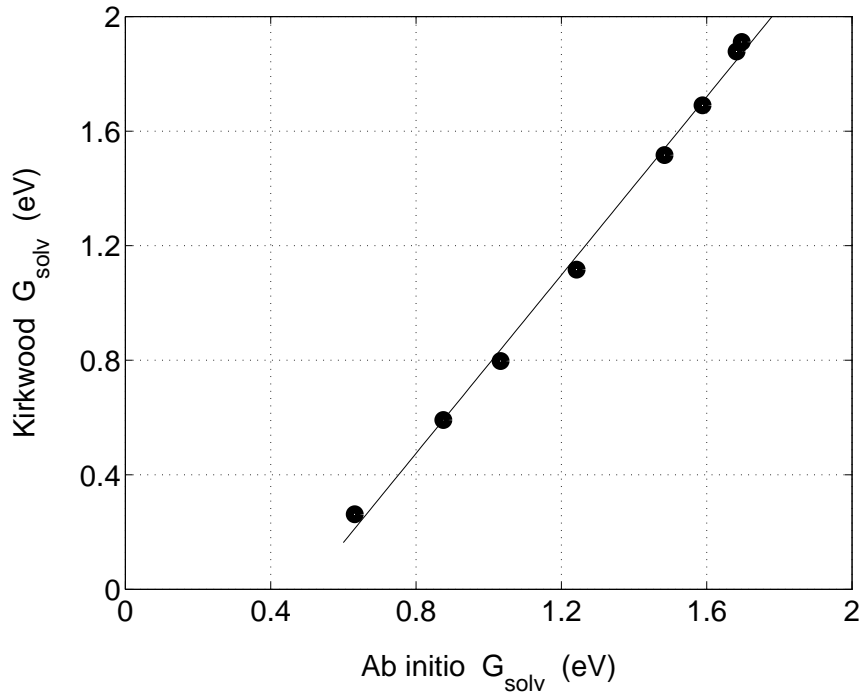}}
\end{center}
\caption[{\em Ab initio} versus Kirkwood solvation energies]
{Correlation plot of the absolute electrostatic solvation free
energies $G^i_{solv}$ as predicted by the {\em a priori} model
(horizontal, $x$) and the Kirkwood model (vertical, $y$) for the
configurations considered in this work (circles). The line is the
least-squares fit, $y = 1.56x - 0.77$ eV.}
\label{fig:abikirkcorr}
\end{figure}

Figure \ref{fig:abikirkcorr} is a correlation plot of the Kirkwood
solvation free energies versus solvation free energies computed with
our {\em a priori} solvation energies.  The most striking feature of
the plot is its linearity.  The fact that there is a vertical
intercept of approximately -0.77~eV is not relevant for the
calculation of free energy differences.  The linear correlation
suggests that although there are quantitative errors, the simplified
Kirkwood model is correct in its qualitative description of solvation
effects and therefore useful for predicting trends in solvation
behavior.

Given the simple form of the dipole term in Eq.~(\ref{Kirkformula}),
and the remarkable linearity of the correlation in Figure
\ref{fig:abikirkcorr}, we could adjust $R$ to modify the scale of the
of the solvation energies and thereby create excellent {\em a
posteriori} agreement with the {\em a priori} results for this one
particular reaction.  Of course, we do not expect such detailed
adjustments to be transferable to other chemical reactions.

\subsection{Comparison of dielectric models}
\label{sec:comparedielec}

We now explore the remarkable linear correlation between the simple
Kirkwood model and our detailed calculations.  The two key
simplifications in the Kirkwood approach are the replacement of the
detailed reactant charge distribution by a dipole, and the replacement
of the arbitrarily shaped dielectric cavity by a sphere.

Our calculations provide us with detailed information regarding the
reactant charge density $\rho_r$, the dielectric function $\epsilon$,
and the induced solvent potential $\phi_s$ for each configuration.
Therefore, we are poised to study the above approximations, which are
not unique to the Kirkwood model but play important roles in a variety
of solvation models employed in the literature \cite{ChemRev}.
In going from the {\em a priori} to the Kirkwood model,
there are two intermediate models, for a total of four models to consider:

\begin{table}[t!]
\begin{center}
\begin{tabular}{lc|cr}
(1) {\em A priori} & & & Dipole approximation (2)\\
& $\epsilon$ = {\em a priori} & $\epsilon$ = {\em a priori} & \\
& $\rho_r$ = {\em a priori}\ \ \ \ \ \ \ & $\rho_r$ = dipole & \\
& $\phi_s$ = {\em a priori} (SC)\ \ \ \ \ \ \ \ & \ \ \ \ \ \
$\phi_s$ = {\em a priori} (NSC) & \\ 
& & & \\
& $\Delta G_1 = \Delta G$ & $\Delta G_2 = 1.15\Delta G$ \\
& & & \\
\hline 
& & &  \\
& $\Delta G_4 = 1.56 \Delta G$ & $\Delta G_3 = 2.28 \Delta G$ \\
& & & \\
& $\epsilon$ = sphere & $\epsilon$ = {\em a priori} & \\
& $\rho_r$ = dipole\ \ \ \ \ \ \ & $\rho_r$ = dipole & \\
& $\phi_s$ = dipole (SC) & \ \ \ \ \ \ $\phi_s$ = dipole
(SC) & \\
(4) Kirkwood model & & & Self-consistent dipole (3) \\
\end{tabular}
\end{center}
\caption[Comparison of dielectric models] {Comparison of the four
dielectric models. $\epsilon$ indicates the shape of the dielectric
cavity, $\rho_r$ indicates the reactant charge density, and $\phi_s$
indicates the induced solvent potential.  SC indicates that $\phi_s$
is determined self-consistently, and NSC, the contrary.  The table
lists solvation free energy differences as calculated in the various
models.}
\label{table:comparemodels}
\end{table}

\begin{enumerate}
\item
Our {\em a priori} approach --- Here, we use the {\em ab initio}
reactant charge density and the {\em a priori} dielectric cavity and
solve the the Poisson equation to obtain the solvation free energy.
\item
Dipole approximation to molecular charge --- Here, we replace the
reactant charge density by a dipole, while holding the solvent
potential $\phi_s$ fixed at its {\em a priori} value from model 1.
Going from model 1 to 2 gauges the effect of replacing the charge
density by a dipole.
\item 
Self-consistent dipole approximation --- Here, we use the dipole of
model 2 and the {\em a priori} dielectric cavity while solving the
Poisson equation self-consistently.  Going from model 2 to 3 gauges
the importance of self-consistency.
\item
Kirkwood model --- Finally, we change the cavity shape to a sphere.
Going from model 3 to 4 gauges the effect of cavity shape.
\end{enumerate}

Table \ref{table:comparemodels} summarizes the above models.  Upon
comparing their predictions for the solvation free energies, we find
the same strong linear correlations we found earlier when comparing
the {\em a priori} and Kirkwood models
(c.f. Figure~\ref{fig:abikirkcorr} in the previous section).  Since
only free energy differences interest us, the slopes of the lines of
best fit are the relevant parameters for comparing the models, and
Table \ref{table:comparemodels} lists these as well.

Upon examining the table, we observe that the effect on energy
differences of replacing the detailed {\em ab initio} charge density
by its dipole moment is only about 15\% (1$\rightarrow$2); the most
important effect is that of self-consistency, resulting in changes of
$\approx$100\% (2$\rightarrow$3); and the impact of the spherical
approximation, resulting in changes of $\approx$50\%, tends to cancel
the preceding effects (3$\rightarrow$4).

These observations lead us to the following conclusion.  Replacing the
{\em ab initio} charge density by its dipole in a realistic dielectric
cavity produces rather large errors in free energies.  Surprisingly,
most of this error arises not from the simplification of the charge
density (a 15\% effect, 1$\rightarrow$2), but rather from the effects
of self-consistency between the reactant charges and the charges
induced in the dielectric cavity (2$\rightarrow$3).  In particular,
the {\em a priori} cavity seems to have an inappropriate shape for use
with a dipole charge distribution as the reduction in error in going
from 3$\rightarrow$4 evinces.  This suggests that simplified charge
densities should be used with correspondingly simplified cavity
shapes.

\section{Conclusions}

We have presented an {\em a priori} approach to the calculation of
solvation free energies using continuum dielectric models coupled to
quantum mechanical calculations.  A derivation of the dielectric
treatment based on a coarse-graining of the molecular description of
the solvent is provided, and this leads to a method for creating the
dielectric cavity which does not rely on empirically scaled van der
Waals radii but rather uses the electron density of the reactants as
the physical variable defining the cavity.  The precise choice of
cavity then is made by ensuring that the dielectric reproduces the
correct linear response of the solvent to electrostatic perturbations.
We study, in a controlled manner, the relative importance of further
approximations that are routinely performed in the literature
including the use of spherical cavities, the replacement of the
reactants by a dipole, or the neglect of self consistency in solving
the electrostatic equation.

Finally, as a model application, we study the hydrolysis of methylene
chloride which has shown unusual solvation effects close to
criticality in super critical oxidation experiments.  Using our {\em a
priori} methodology, results in good agreement with available
experimental reaction barriers are found which explain these unusual
trends.

\section{Acknowledgements}

This work was supported primarily by the MRSEC Program of the National
Science Foundation under award number DMR 94-00334 and also by the
Alfred P. Sloan Foundation (BR-3456).  This work made use of the
Cornell Center for Materials Research Shared Experimental Facilities,
supported through the NSF MRSEC program (DMR-9632275).  Dielectric
calculations were carried out on the MIT Xolas prototype SMP cluster.

\bibliography{mybib}
\bibliographystyle{plain}

\end{document}